\documentclass{article}

\usepackage{epsf,multicol,ifthen}
\usepackage[cp1251]{inputenc}
\usepackage[dvips]{graphicx}
\usepackage{cite}
\usepackage{hyperref}

\title{Multipolar model of bremsstrahlung accompanying proton-decay of nuclei}

\author{Sergei~P.~Maydanyuk\thanks{E-mail: maidan@kinr.kiev.ua} \\
\small\emph{Institute for Nuclear Research, National Academy of Science of Ukraine} \\
\small\emph{47, prosp. Nauki, Kiev, 03680, Ukraine}}

\date{\small\today}

\begin{document}

\maketitle

\begin{abstract}
Emission of bremsstrahlung photons accompanying proton decay of nuclei is studied. The new improved multipolar model describing such a process is presented. The angular formalism of calculations of the matrix elements is stated in details. The bremsstrahlung probabilities for the $^{157}{\rm Ta}$, $^{161}{\rm Re}$, $^{167}{\rm Ir}$ and $^{185}{\rm Bi}$ nuclei decaying from the $2s_{1/2}$ state,
the $^{109}_{53}{\rm I}_{56}$ and $^{112}_{55}{\rm Cs}_{57}$ nuclei decaying from the $1d_{5/2}$ state,
the $^{146}_{69}{\rm Tm}_{77}$ and $^{151}_{71}{\rm Lu}_{80}$ nuclei decaying from the $0h_{11/2}$ state are predicted.
Such spectra have orders of values similar to the experimental data for the bremsstrahlung photons emitted during the $\alpha$-decay. This indicates on real possibility to study bremsstrahlung photons during proton decay experimentally and perform further measurements.
\end{abstract}


\vspace{0mm}

\section{Introduction
\label{introduction}}

Last two decades many experimental and theoretical efforts have been made to investigate the nature of the bremsstrahlung emission accompanying $\alpha$-decay of heavy nuclei
\cite{Batkin.1986.SJNCA,D'Arrigo.1994.PHLTA,Dyakonov.1996.PRLTA,
Kasagi.1997.JPHGB,
Kasagi.1997.PRLTA,
Papenbrock.1998.PRLTA,
Dyakonov.1999.PHRVA,
Bertulani.1999.PHRVA,
Takigawa.1999.PHRVA,
Flambaum.1999.PRLTA,
Tkalya.1999.JETP,
Tkalya.1999.PHRVA,
Misicu.2001.JPHGB,
Dijk.2003.FBSSE,
Maydanyuk.2003.PTP,
Maydanyuk.2006.EPJA,
Ohtsuki.2006.CzJP,
Amusia.2007.JETP,
Boie.2007.PRL,
Jentschura.2008.PRC,
Maydanyuk.2008.EPJA,
Giardina.2008.MPLA}
and spontaneous fission
\cite{Smith.1956.PR,Milton.1958.PR,Verbinski.1973.PRC,Brooks.1973.PRC,Sobel.1973.PRC,Dietrich.1974.PRC,
Kasagi.1989.JPSJ,Luke.1991.PRC,Hofman.1993.PRC,Ploeg.1995.PRC,
Varlachev.2007.IRAN,Eremin.2010.IJMPE,Maydanyuk.2010.PRC,Pandit.2010.PLB}.
A key idea of such researches consists in finding a new method of extraction of a new information about dynamics of the studied nuclear processes from measured bremsstrahlung spectra.
Parallel study of the interested type of nuclear decay and the bremsstrahlung photons which are emitted during it, gives us more complete, richer information about the studied nuclear process.
As examples, we established dependence between nuclear deformation and the bremsstrahlung spectrum accompanying $\alpha$-decay of the $^{226}{\rm Ra}$ nucleus \cite{Maydanyuk.2009.NPA}.
Analysis of the bremsstrahlung emission which accompanies ternary fission of $^{252}{\rm Cf}$, allow us to understand better geometry of undergoing fission process in its first stage \cite{Maydanyuk.2010.JPCS}.


In our previous paper \cite{Giardina.2008.MPLA} we compared the bremsstrahlung spectra in $\alpha$-decay of the $^{214}{\rm Po}$ and $^{226}{\rm Ra}$ nuclei, which have similar daughter nucleus--$\alpha$ particle potentials, and noted a clear difference between these spectra. In the cited paper we concluded that the different slopes of spectra were connected with the different $Q$-values of $\alpha$-decay for two considered nuclei ($Q_{\alpha}=7.83$ MeV for $^{214}{\rm Po}$ and $Q_{\alpha}=4.87$ MeV for $^{226}{\rm Ra}$) and we confirmed that such difference gives different contributions
of emission from the tunnelling region into the total spectra.
Now I put a question: which other parameters have important influence on the bremsstrahlung spectrum?

Let us consider Figs.~4 and 7 in Ref.~\cite{Ploeg.1995.PRC} where the calculated $\gamma$-ray emission probabilities for the spontaneous fission of $^{252}{\rm Cf}$ are presented. One can see that the calculated spectrum is changed in dependence on mass split. So, one can suppose that the emission probability should be dependent on numbers of masses and charges of the daughter nucleus and the emitted fragment. The idea in Ref.~\cite{So_Kim.2000.JKPS} about influence of the electromagnetic charge of the daughter nucleus on the bremsstrahlung probability reflects this property only partially while \emph{effective charge of the decaying system is connected with such property more directly}. Now if to consider the formula of the bremsstrahlung probability (for example, see (1) in \cite{Maydanyuk.2006.EPJA}) then one can find its direct dependence on square of the effective charge, $Z_{\rm eff}^{2}$, i.~e. we obtain a real basis for such supposition.

As another type of decay which has larger effective charge in comparison with $\alpha$-decay, the emission of proton from nucleus can be analyzed (for example, $Z_{\rm eff}^{2}$ equals to 0.286, 0.285, 0.288 and 0.303 for the proton emitters $^{157}{\rm Ta}$, $^{161}{\rm Re}$, $^{167}{\rm Ir}$ and $^{185}{\rm Bi}$, while it equals to 0.185 and 0.16 for the $\alpha$-decaying nuclei $^{214}{\rm Po}$ and $^{210}{\rm Po}$, correspondingly).
The first indication on possibility of presence of such process was given in \cite{Kurgalin.2001.IRAN} where authors estimated the averaged over all angles spectrum for the $^{113}{\rm Cs}$ nucleus up to 250~keV of energy of the emitted photons.
However, the detained study of such process has never been performed.
First of all, a question is appearing whether the emission of photons accompanying such type of decay, is enough intensive to measure it experimentally ($Q_{p}$-values for proton decay are smaller than for $\alpha$-decay)?
We are able to calculate the bremsstrahlung probabilities for the $\alpha$-decays, which turn out to be in enough well agreement with experimental data, without normalization of the theoretical spectra on experiment (see \cite{Maydanyuk.2009.TONPPJ,Maydanyuk.2009.JPS}).
Used in such calculations, the multipole approach (started from \cite{Tkalya.1999.JETP,Maydanyuk.2003.PTP}) looks to be the most accurate and corrected in angular description of the photons emission during the $\alpha$-decay. It turns out that such model allows to calculate absolute values of the probability without any normalization relatively experimental data, and achieves enough good agreement with them.
So, it could be a convenient basis for estimation of the probabilities of the bremsstrahlung photons in the proton-decay.
According to our analysis, the dipole approximation of the wave function of photons (started from \cite{Papenbrock.1998.PRLTA}) gives the overvalued spectra in absolute scale (see Fig.~9 in~\cite{Maydanyuk.arXiv.0904.2247}), and by such a reason we shall not use it in this paper.

Majority of nuclei emit protons in the state with nonzero orbital quantum number. So, from point of view of improvement of theory, it could be useful to generalize formalism of determination of matrix element of emission, taking into account such states. Inclusion of the states with nonzero orbital moment requires to add spin-orbital component into potential of interaction between the daughter nucleus and proton emitted (never considered before in this research). Note that up to-date it has not been known anything about the emission of photons from such states. So, it could be interesting to perform such calculations, that causes necessity of further improvement of theory of emission of bremsstrahlung photons in decays of nuclei.

This paper answers on these questions, which is organized so. At first, the improved multipole model of the bremsstrahlung photons emitted during proton decay is presented, where emphasis is made on construction of the angular formalism of the matrix elements and calculation of the absolute bremsstrahlung probability.
On its basis I perform theoretical study of the bremsstrahlung emission for some proton emitters, give predictions (in absolute scale) and analyze them.


\section{Model
\label{sec.2}}

\subsection{Matrix element of transition
\label{sec.2.1}}

I define the matrix element like (2.11) in Ref.~\cite{Maydanyuk.2003.PTP} (in the first correction of the non-stationary perturbation theory with stationary limits $t_{0}=-\infty$ and $t_{1}=+\infty$,
and with normalization $|C| \to 1$):
\begin{equation}
  a_{fi} = F_{fi} \cdot 2\pi \,\delta(w_{f}-w_{i}+w),
\label{eq.2.1.1}
\end{equation}
where
\begin{equation}
\begin{array}{lcl}
  \vspace{2mm}
  F_{fi} & = &
    Z_{eff}\, \displaystyle\frac{e}{m} \,
    \sqrt{\displaystyle\frac{2\pi\hbar}{w}} \cdot p\,(k_{i},k_{f}), \\

  \vspace{2mm}
  p\,(k_{i}, k_{f}) & = &
    \displaystyle\sum\limits_{\alpha=1,2} \mathbf{e}^{(\alpha),*}\, \mathbf{p}\,(k_{i}, k_{f}), \\

  \vspace{2mm}
  \mathbf{p}\,(k_{i}, k_{f}) & = &
    \biggl< k_{f} \biggl| \,  e^{-i\mathbf{kr}} \displaystyle\frac{\partial}{\partial \mathbf{r}} \,
    \biggr| \,k_{i} \biggr> =
    \int
      \psi^{*}_{f}(\mathbf{r}) \:
      e^{-i\mathbf{kr}} \displaystyle\frac{\partial}{\partial \mathbf{r}}\:
      \psi_{i}(\mathbf{r}) \;
      \mathbf{dr}
\end{array}
\label{eq.2.1.2}
\end{equation}
and $\psi_{i}(\mathbf{r}) = |k_{i}\bigr>$ and $\psi_{f}(\mathbf{r}) = |k_{f}\bigr>$ are stationary wave functions of the decaying system in the initial $i$-state and final $f$-state which do not contain number of photons emitted,
$Z_{\rm eff}$ and $m$ are effective charge and reduced mass of this system.
$\mathbf{e}^{(\alpha)}$ are unit vectors of polarization of the photon emitted, $\mathbf{k}$ is wave vector of the photon and $w = k = \bigl| \mathbf{k}\bigr|$. Vectors $\mathbf{e}^{(\alpha)}$ are perpendicular to $\mathbf{k}$ in Coulomb calibration. We have two independent polarizations $\mathbf{e}^{(1)}$ and $\mathbf{e}^{(2)}$ for the photon with impulse $\mathbf{k}$ ($\alpha=1,2$).
One can develop formalism simpler in the system of units where $\hbar = 1$ and $c = 1$, but we shall write constants $\hbar$ and $c$ explicitly.
Let us find also square of the matrix element $a_{fi}$ used in definition of \emph{probability of transition}.
Using the \emph{formula of power reduction of $\delta$-function} (see~\cite{Bogoliubov.1980}, \S~21, p.~169):
\begin{equation}
  [\delta(w)]^{2} = \delta(w)\: \delta(0) = \delta(w) \: (2\pi)^{-1} \int dt =
  \delta(w)\: (2\pi)^{-1}\, T,
\label{eq.2.1.3}
\end{equation}
we find ($T\to +\infty$ is higher time limit):
\begin{equation}
  |a_{fi}|^{2} = 2\pi\: T\: |F_{fi}|^{2} \cdot \delta(w_{f}-w_{i}+w),
\label{eq.2.1.4}
\end{equation}
that looks like (4.21) in Ref.~\cite{Bogoliubov.1980} (with accuracy up to factor $(2\pi)^{2}$) and
like (42.5) in Ref.~\cite{Landau.v3.1989} (exactly, see \S~42, p.~189).


\subsection{Linear and circular polarizations of the photon emitted
\label{sec.2.2}}

Rewrite vectors of \emph{linear} polarization $\mathbf{e}^{(\alpha)}$ through \emph{vectors of circular polarization} $\mathbf{\xi}_{\mu}$ with opposite directions of rotation (see Ref.~\cite{Eisenberg.1973}, (2.39), p.~42):
\begin{equation}
\begin{array}{ccc}
  \mathbf{\xi}_{-1} = \displaystyle\frac{1}{\sqrt{2}}\,
                      \bigl(\mathbf{e}^{(1)} - i\mathbf{e}^{(2)}\bigr), &
  \mathbf{\xi}_{+1} = -\displaystyle\frac{1}{\sqrt{2}}\,
                      \bigl(\mathbf{e}^{(1)} + i\mathbf{e}^{(2)}\bigr), &
  \mathbf{\xi}_{0} = \mathbf{e}^{(3)} = 0.
\end{array}
\label{eq.2.2.1}
\end{equation}
Then $p\,(k_{i},k_{f})$ can be rewritten so:
\begin{equation}
  p\,(k_{i}, k_{f}) =
    \sum\limits_{\mu = -1, 1}  h_{\mu}\,\mathbf{\xi}^{*}_{\mu}
    \int
      \psi^{*}_{f}(\mathbf{r})\:
      e^{-i\mathbf{kr}} \displaystyle\frac{\partial}{\partial \mathbf{r}} \:
      \psi_{i}(\mathbf{r}) \;
    \mathbf{dr},
\label{eq.2.2.2}
\end{equation}
\begin{equation}
\begin{array}{ccc}
  h_{\pm} = \mp \displaystyle\frac{1 \pm i}{\sqrt{2}}, &
  h_{-1} + h_{+1} = -i\sqrt{2}, &
  \sum\limits_{\alpha = 1,2} \mathbf{e}^{(\alpha),*} =
    h_{-1} \mathbf{\xi}_{-1}^{*} + h_{+1} \mathbf{\xi}_{+1}^{*}.
\end{array}
\label{eq.2.2.3}
\end{equation}


\subsection{Expansion of the vector potential $\mathbf{A}$ by multipoles
\label{sec.2.3}}

In further calculations of the matrix element $p\,(k_{i}, k_{f})$ the different expansions of function $e^{-i\mathbf{kr}}$ connected with the vector potential $\mathbf{A}$ of the electro-magnetic field of the daughter nucleus can be used. In spherically symmetric approximation
I shall use the multipole expansion
(see Ref.~\cite{Eisenberg.1973}, (2.106) in p.~58):
\begin{equation}
  \mathbf{\xi}_{\mu}\, e^{i \mathbf{kr}} =
    \mu\, \sqrt{2\pi}\, \sum_{l}\,
    (2l+1)^{1/2}\, i^{l}\,  \cdot
    \Bigl[ \mathbf{A}_{l\mu} (\mathbf{r}, M) +
    i\mu\, \mathbf{A}_{l\mu} (\mathbf{r}, E) \Bigr],
\label{eq.2.3.1}
\end{equation}
where (see Ref.~\cite{Eisenberg.1973}, (2.73) in p.~49, (2.80) in p.~51)
\begin{equation}
\begin{array}{lcl}
  \vspace{2mm}
  \mathbf{A}_{l\mu}(\mathbf{r}, M) & = &
        j_{l}(kr) \: \mathbf{T}_{ll,\mu} ({\mathbf n}_{ph}), \\
  \vspace{2mm}
  \mathbf{A}_{l\mu}(\mathbf{r}, E) & = &
        \sqrt{\displaystyle\frac{l+1}{2l+1}}\,
        j_{l-1}(kr) \: \mathbf{T}_{ll-1,\mu}({\mathbf n}_{ph})\; - \\
  & - &
        \sqrt{\displaystyle\frac{l}{2l+1}}\,
        j_{l+1}(kr) \: \mathbf{T}_{ll+1,\mu}({\mathbf n}_{ph}).
\end{array}
\label{eq.2.3.2}
\end{equation}
Here, $\mathbf{A}_{l\mu}(\textbf{r}, M)$ and $\mathbf{A}_{l\mu}(\textbf{r}, E)$ are \emph{magnetic} and \emph{electric multipoles}, $j_{l}(kr)$ are \emph{spherical Bessel functions of order $l$}, $\mathbf{T}_{ll',\mu}(\mathbf{n})$ are \emph{vector spherical harmonics}.
We orientate the frame system so that axis $z$ is parallel to the vector $\mathbf{k}$.
The functions $\mathbf{T}_{ll',\mu}(\mathbf{n})$ have the following form
(${\mathbf \xi}_{0} = 0$, see Ref.~\cite{Eisenberg.1973}, p.~45):
\begin{equation}
  \mathbf{T}_{jl,m} (\mathbf{n}) =
  \sum\limits_{\mu = \pm 1} (l, 1, j \,\big| \,m-\mu, \mu, m) \; Y_{l,m-\mu}(\mathbf{n}) \; \mathbf{\xi}_{\mu},
\label{eq.2.3.3}
\end{equation}
where $(l, 1, j \,\bigl| \, m-\mu, \mu, m)$ are \emph{Clebsh-Gordon coefficients} and
$Y_{lm}(\theta, \varphi)$ are \emph{spherical functions} defined, according to~\cite{Landau.v3.1989} (see p.~119, (28,7)--(28,8)).



In the spherically symmetric approximation, wave functions of the decaying system in the initial and final states are separated into the radial and angular components, and these states are characterized by quantum numbers $l$ and $m$. We shall be interesting in such photon emission when the system transits to superposition of all possible final states with different magnetic numbers $m$ at the same orbital number $l$ (for both states).
So, let's write wave functions as
\begin{equation}
\begin{array}{lcl}
  \psi_{i} (\mathbf{r}, l_{i}) & = &
    \varphi_{i} (r, l_{i}) \: \displaystyle\sum\limits_{m_{i}} Y_{l_{i}m_{i}}({\mathbf n}_{\rm r}^{i}), \\
  \psi_{f} (\mathbf{r}, l_{f}) & = &
    \varphi_{f} (r, l_{f}) \: \displaystyle\sum\limits_{m_{f}} Y_{l_{f}m_{f}}({\mathbf n}_{\rm r}^{f})
\end{array}
\label{eq.2.3.4}
\end{equation}
and obtain:
\begin{equation}
  p\,(k_{i}, k_{f}) = \sqrt{2\pi}\: \sum\limits_{l} \: (-i)^{l}\, \sqrt{2l+1} \: \Bigl[ p_{l}^{M} -ip_{l}^{E} \Bigr],
\label{eq.2.3.5}
\end{equation}
where
\begin{equation}
\begin{array}{ll}
  p_{l}^{M} = \sum\limits_{\mu = -1, 1} \mu\, h_{\mu}\: p_{l\mu}^{M}, &
  p_{l}^{E} = \sum\limits_{\mu = -1, 1} \mu^{2} h_{\mu}\: p_{l\mu}^{E}
\end{array}
\label{eq.2.3.6}
\end{equation}
and
\begin{equation}
\begin{array}{lcl}
  p_{l\mu}^{M} & = &
        \displaystyle\int\limits^{+\infty}_{0} dr
        \displaystyle\int d\Omega \: r^{2} \,
        \psi^{*}_{f}(\mathbf{r}) \,
        \biggl( \displaystyle\frac{\partial}{\partial \mathbf{r}}\, \psi_{i}(\mathbf{r}) \biggr) \,
        \mathbf{A}_{l\mu}^{*} (\mathbf{r}, M), \\

  p_{l\mu}^{E} & = &
        \displaystyle\int\limits^{+\infty}_{0} dr
        \displaystyle\int d\Omega \: r^{2} \,
        \psi^{*}_{f}(\mathbf{r}) \,
        \biggl( \displaystyle\frac{\partial}{\partial \mathbf{r}}\, \psi_{i}(\mathbf{r}) \biggr)\,
        \mathbf{A}_{l\mu}^{*} (\mathbf{r}, E).
\end{array}
\label{eq.2.3.7}
\end{equation}

Using \emph{gradient formula} (see (2.56), p.~46 in~\cite{Eisenberg.1973}):
\begin{equation}
\begin{array}{lcl}
  \displaystyle\frac{\partial}{\partial \mathbf{r}}\: \psi_{i}(\mathbf{r}) & = &
  \displaystyle\frac{\partial}{\partial \mathbf{r}}\:
    \Bigl\{ \varphi_{i} (r)\: \displaystyle\sum\limits_{m_{i}} Y_{l_{i}m_{i}}({\mathbf n}_{r}^{i}) \Bigr\} = \\

  & = &
    \sqrt{\displaystyle\frac{l_{i}}{2l_{i}+1}}\:
    \biggl( \displaystyle\frac{d\varphi_{i}(r)}{dr} + \displaystyle\frac{l_{i}+1}{r}\, \varphi_{i}(r) \biggr)\,
    \displaystyle\sum\limits_{m_{i}}
      \mathbf{T}_{l_{i} l_{i}-1, m_{i}}({\mathbf n}_{r}^{i}) - \\
  & - &
  \sqrt{\displaystyle\frac{l_{i}+1}{2l_{i}+1}}\:
    \biggl( \displaystyle\frac{d\varphi_{i}(r)}{dr} - \displaystyle\frac{l_{i}}{r}\, \varphi_{i}(r) \biggr)\,
    \displaystyle\sum\limits_{m_{i}}
      \mathbf{T}_{l_{i} l_{i}+1, m_{i}}({\mathbf n}_{r}^{i}),
\end{array}
\label{eq.2.3.8}
\end{equation}
we obtain:
\begin{equation}
\begin{array}{lcl}
\vspace{1mm}
  p_{l_{ph}}^{M} & = &
    \sqrt{\displaystyle\frac{l_{i}}{2l_{i}+1}}\:
      I_{M}\,(l_{i}, l_{f}, l_{ph}, l_{i}-1) \times \\
\vspace{1mm}
  & \times &
      \Bigl\{
        J_{1}(l_{i},l_{f},l_{\rm ph}) +
        (l_{i}+1) \cdot J_{2}(l_{i},l_{f},l_{\rm ph})
      \Bigr\}\; - \\
\vspace{3mm}
  & - &
    \sqrt{\displaystyle\frac{l_{i}+1}{2l_{i}+1}}\:
      I_{M}\, (l_{i}, l_{f}, l_{ph}, l_{i}+1) \cdot
      \Bigl\{J_{1}(l_{i},l_{f},l_{\rm ph}) - l_{i} \cdot J_{2}(l_{i},l_{f},l_{\rm ph}) \Bigr\},
\end{array}
\label{eq.2.3.9}
\end{equation}
\begin{equation}
\begin{array}{lcl}
\vspace{1mm}
  p_{l_{ph}}^{E} & = &
    \sqrt{\displaystyle\frac{l_{i}\,(l_{\rm ph}+1)}{(2l_{i}+1)(2l_{\rm ph}+1)}} \cdot
      I_{E}\,(l_{i}, l_{f}, l_{ph}, l_{i}-1, l_{\rm ph}-1) \times \\
\vspace{1mm}
  & \times &
      \Bigl\{
        J_{1}(l_{i},l_{f},l_{\rm ph}-1)\; +
        (l_{i}+1) \cdot J_{2}(l_{i},l_{f},l_{\rm ph}-1)
      \Bigr\}\; - \\
\vspace{1mm}
    & - &
    \sqrt{\displaystyle\frac{l_{i}\,l_{\rm ph}}{(2l_{i}+1)(2l_{\rm ph}+1)}} \cdot
      I_{E}\,(l_{i}, l_{f}, l_{ph}, l_{i}-1, l_{\rm ph}+1) \times \\
\vspace{1mm}
  & \times &
      \Bigl\{
        J_{1}(l_{i},l_{f},l_{\rm ph}+1)\; +
        (l_{i}+1) \cdot J_{2}(l_{i},l_{f},l_{\rm ph}+1)
      \Bigr\}\; + \\
\vspace{1mm}
  & + &
    \sqrt{\displaystyle\frac{(l_{i}+1)(l_{\rm ph}+1)}{(2l_{i}+1)(2l_{\rm ph}+1)}} \cdot
      I_{E}\,(l_{i}, l_{f}, l_{ph}, l_{i}+1, l_{\rm ph}-1) \times \\
\vspace{1mm}
  & \times &
      \Bigl\{
        J_{1}(l_{i},l_{f},l_{\rm ph}-1)\; -
        l_{i} \cdot J_{2}(l_{i},l_{f},l_{\rm ph}-1)
      \Bigr\}\; - \\
  & - &
    \sqrt{\displaystyle\frac{(l_{i}+1)\,l_{\rm ph}}{(2l_{i}+1)(2l_{\rm ph}+1)}} \cdot
      I_{E}\,(l_{i}, l_{f}, l_{ph}, l_{i}+1, l_{\rm ph}+1) \times \\
\vspace{1mm}
  & \times &
      \Bigl\{
        J_{1}(l_{i},l_{f},l_{\rm ph}+1)\; -
        l_{i} \cdot J_{2}(l_{i},l_{f},l_{\rm ph}+1)
      \Bigr\},
\end{array}
\label{eq.2.3.10}
\end{equation}
where
\begin{equation}
\begin{array}{ccl}
  J_{1}(l_{i},l_{f},n) & = &
  \displaystyle\int\limits^{+\infty}_{0}
    \varphi^{*}_{f}(l_{f},r)\, \displaystyle\frac{d\varphi_{i}(r, l_{i})}{dr}\,
    j_{n}(kr)\; r^{2} dr, \\

  J_{2}(l_{i},l_{f},n) & = &
  \displaystyle\int\limits^{+\infty}_{0}
    \varphi^{*}_{f}(l_{f},r)\, \varphi_{i}(r, l_{i})\: j_{n}(kr)\; r\, dr, \\
\end{array}
\label{eq.2.3.11}
\end{equation}
\begin{equation}
\begin{array}{ccl}
  I_{M}\, (l_{i}, l_{f}, l_{\rm ph}, l_{1}) & = &
    \displaystyle\sum\limits_{\mu = \pm 1}
    \displaystyle\sum\limits_{m_{i}}
    \displaystyle\sum\limits_{m_{f}}
      \mu h_{\mu}
    \displaystyle\int
      Y_{l_{f}m_{f}}^{*}({\mathbf n}_{\rm r}^{f})\,
      \mathbf{T}_{l_{i}\, l_{1},\, m_{i}}(\mathbf{n}^{i}_{\rm r})\,
      \mathbf{T}_{l_{\rm ph}\,l_{\rm ph},\, \mu}^{*}({\mathbf n}_{\rm ph})\; d\Omega, \\

  I_{E}\, (l_{i}, l_{f}, l_{\rm ph}, l_{1}, l_{2}) & = &
    \displaystyle\sum\limits_{\mu = \pm 1}
    \displaystyle\sum\limits_{m_{i}}
    \displaystyle\sum\limits_{m_{f}}
      h_{\mu}
    \displaystyle\int
      Y_{l_{f}m_{f}}^{*}({\mathbf n}_{\rm r}^{f})\,
      \mathbf{T}_{l_{i} l_{1},\, m_{i}}(\mathbf{n}^{i}_{\rm r})\,
      \mathbf{T}_{l_{\rm ph} l_{2},\, \mu}^{*}({\mathbf n}_{\rm ph})\; d\Omega.
\end{array}
\label{eq.2.3.12}
\end{equation}

For the case $l_{i}=0$ we have essentially simpler formulas. In particular, for the matrix components we obtain:
\begin{equation}
\begin{array}{lcl}
  \vspace{2mm}
  p_{l_{ph}}^{M} & = & -\; I_{M}(0,l_{f},l_{ph}, l_{ph}) \cdot J\,(l_{f},l_{\rm ph}), \\
  \vspace{2mm}
  p_{l_{ph}}^{E} & = &
    -\sqrt{\displaystyle\frac{l_{ph}+1}{2l_{ph}+1}}\; I_{E}(0,l_{f},l_{ph},1,l_{ph}-1) \cdot
    J\, (l_{f},l_{\rm ph}-1)\; + \\
  & + &
    \sqrt{\displaystyle\frac{l_{ph}}{2l_{ph}+1}}\; I_{E}(0,l_{f},l_{ph},1,l_{ph}+1)
    \cdot J\, (l_{f},l_{\rm ph}+1),
\end{array}
\label{eq.2.3.13}
\end{equation}
where we use the simplified notation:
\begin{equation}
  J\,(l_{f}, n) = J_{1}\,(0, l_{f}, n).
\label{eq.2.3.14}
\end{equation}
Using the following value of the Clebsh-Gordon coefficient: 
\begin{equation}
  (110\,|1, -1, 0) = (110\,|-1, 1, 0) = \sqrt{\displaystyle\frac{1}{3}},
\label{eq.2.3.15}
\end{equation}
from (\ref{eq.2.3.3}) and (\ref{eq.2.3.15}) we obtain:
\begin{equation}
\begin{array}{c}
  \mathbf{T}_{01,0}(\mathbf{n}^{i}_{r}) =
    \displaystyle\sum\limits_{\mu = \pm 1} (110\,|-\mu\mu 0) \:
      Y_{1,-\mu}(\mathbf{n}^{i}_{r}) \: \mathbf{\xi}_{\mu} =
    \sqrt{\displaystyle\frac{1}{3}}
      \displaystyle\sum\limits_{\mu = \pm 1}
      Y_{1,-\mu}(\mathbf{n}^{i}_{r}) \: \mathbf{\xi}_{\mu}
\end{array}
\label{eq.2.3.16}
\end{equation}
and for the angular integrals for transition into the superposition of all possible final $f$-states
with different $m_{f}$ at the same $l_{f}$ from eq.~(\ref{eq.2.3.12}) we obtain:
\begin{equation}
\begin{array}{lcl}
  I_{M}(0,l_{f},l_{ph},n) & = &
    \sqrt{\displaystyle\frac{1}{3}}
    \displaystyle\sum\limits_{\mu = \pm 1} \mu\, h_{\mu}
    \sum\limits_{\mu^{\prime} = \pm 1} (n, 1, l_{ph} \big|\, \mu-\mu^{\prime}, \mu^{\prime}, \mu)\; \times \\
  & \times &
    \displaystyle\int \:
      Y_{l_{f}m}^{*}({\mathbf n}_{r}^{f}) \,
      Y_{1,-\mu^{\prime}}(\mathbf{n}_{r}^{i}) \,
      Y_{n, \mu-\mu^{\prime}}^{*}(\mathbf{n}_{ph}) \;
      d\Omega, \\

  I_{E}(0,l_{f},l_{ph},1,n) & = &
    \sqrt{\displaystyle\frac{1}{3}}
    \displaystyle\sum\limits_{\mu = \pm 1} h_{\mu}
    \sum\limits_{\mu^{\prime} = \pm 1} (n, 1, l_{ph} \big|\, \mu-\mu^{\prime}, \mu^{\prime}, \mu)\; \times \\
  & \times &
    \displaystyle\int \:
      Y_{l_{f}m}^{*}({\mathbf n}_{r}^{f}) \,
      Y_{1,-\mu^{\prime}}(\mathbf{n}_{r}^{i}) \,
      Y_{n, \mu-\mu^{\prime}}^{*}(\mathbf{n}_{ph}) \;
      d\Omega.
\end{array}
\label{eq.2.3.17}
\end{equation}

\subsection{Vectors $\mathbf{n}^{i}_{r}$, $\mathbf{n}^{f}_{r}$, $\mathbf{n}_{ph}$ and calculations of the angular integrals at $l_{i}=0$
\label{sec.2.5}}

Let us analyze physical sense of vectors $\mathbf{n}^{i}_{r}$, $\mathbf{n}^{f}_{r}$ and $\mathbf{n}_{ph}$. According to definition of wave functions $\psi_{i} (\mathbf{r})$ and $\psi_{f} (\mathbf{r})$, the vectors $\mathbf{n}^{i}_{r}$ and $\mathbf{n}^{f}_{r}$ determine orientation of radius-vector $\mathbf{r}$ from the center of frame system to point where this wave functions describes the particle before and after the emission of photon. Such description of the particle has a probabilistic sense and is fulfilled over whole space.
Change of direction of motion (or tunneling) of the particle in result of the photon emission can be characterized by change of quantum numbers $l$ and $m$ in the angular wave function: $Y_{00}(\mathbf{n}^{i}_{r}) \to Y_{lm}(\mathbf{n}^{f}_{r})$ (which changes the probability of appearance of this particle along different directions, and angular asymmetry is appeared).
The vector $\mathbf{n}_{ph}$ determines orientation of radius-vector $\mathbf{r}$ from the center of the frame system to point where wave function of photon describes its ``appearance''. Using such a logic, we have:
\begin{equation}
  \mathbf{n}_{ph} = \mathbf{n}^{i}_{r} = \mathbf{n}^{f}_{r} = \mathbf{n}_{r}.
\label{eq.2.5.1}
\end{equation}
As we use the frame system where axis $z$ is parallel to vector $\mathbf{k}$ of the photon emission, then dependent on $\mathbf{r}$ integrant function in the matrix element represents amplitude (its square is probability) of appearance of the particle at point $\mathbf{r}$ after emission of photon, if this photon has emitted along axis $z$. Then angle $\theta$ (of vector $\mathbf{n}_{\mathbf{r}}$) is the angle between direction of the particle motion (with possible tunneling) and direction of the photon emission.

Let us consider the angular integral in (\ref{eq.2.3.17}) over $d\,\Omega$. Using (\ref{eq.2.5.1}), we find:
\begin{equation}
\begin{array}{l}
  \vspace{2mm}
  \displaystyle\int \:
    Y_{lm}^{*}({\mathbf n}_{r}) \,
    Y_{1,-\mu^{\prime}}(\mathbf{n}_{r}) \,
    Y_{n, \mu-\mu^{\prime}}^{*}(\mathbf{n_{r}}) \;
    d\Omega\; =\;
  (-1)^{l+n-\mu^{\prime}+1 + \frac{|m+\mu^{\prime}|}{2}} \; i^{l+n+1} \; \times \\
  \vspace{2mm}
  \;\times\;
    \sqrt{\displaystyle\frac{3\,(2l+1)\,(2n+1)}{32\pi}\;
          \displaystyle\frac{(l-1)!}{(l+1)!} \;
          \displaystyle\frac{(n-|m+\mu^{\prime}|)!}{(n+|m+\mu^{\prime}|)!}}\;
    \times \\
  \; \times\;
    \displaystyle\int\:
      P_{l}^{1}(\cos{\theta}) \; P_{1}^{1}(\cos{\theta}) \; P_{n}^{|m+\mu^{\prime}|} (\cos{\theta}) \cdot
      \sin{\theta} \, d\theta \,d\varphi,
\end{array}
\label{eq.2.5.2}
\end{equation}
where $P_{l}^{m}(\cos{\theta})$ are \emph{associated Legandre's polynomial} (see \cite{Landau.v3.1989}, p.~752--754, (c,1)--(c,4); also see~\cite{Eisenberg.1973} (2.6), p.~34) and the following restrictions on possible values of $m$ and $l_{f}$
have been obtained:
\begin{equation}
\begin{array}{ccc}
  m = -\mu = \pm 1,  &  l_{f} \ge 1, &
  n \ge |\mu - \mu^{\prime}| = |m + \mu^{\prime}|.
\end{array}
\label{eq.2.5.3}
\end{equation}

Let us introduce the following differential matrix elements $dp_{l}^{M}$ and $dp_{l}^{E}$ dependent on the
angle $\theta$ :
\begin{equation}
\begin{array}{lcl}
  \displaystyle\frac{d \,p_{l}^{M}}{\sin{\theta}\,d\theta} & = &
    i^{l_{f}+l_{ph}+1} \;
    J(l_{f},l_{ph})
    \displaystyle\sum\limits_{m = \pm 1}
    m \,h_{-m} \;
    \displaystyle\sum\limits_{\mu^{\prime} = \pm 1}
    C_{l_{f}l_{ph}l_{ph}}^{m \mu^{\prime}} f_{l_{f}l_{ph}}^{m \mu^{\prime}}(\theta), \\

  \displaystyle\frac{d \,p_{l}^{E}}{\sin{\theta}\,d\theta} & = &
    -i^{l_{f}+l_{ph}} \;
    \sqrt{\displaystyle\frac{l_{ph}+1}{2l_{ph}+1}} \, J(l_{f},l_{ph}-1)\; \times \\
  & \times &
    \displaystyle\sum\limits_{m = \pm 1}
    h_{-m} \;
    \displaystyle\sum\limits_{\mu^{\prime} = \pm 1}
      C_{l_{f},l_{ph},l_{ph}-1}^{m \mu^{\prime}} \: f_{l_{f},l_{ph}-1}^{m \mu^{\prime}}(\theta) \: - \\
  & - &
    i^{l_{f}+l_{ph}} \;
    \sqrt{\displaystyle\frac{l_{ph}}{2l_{ph}+1}} \, J(l_{f},l_{ph}+1)\; \times \\
  & \times &
    \displaystyle\sum\limits_{m = \pm 1}
    h_{-m} \;
    \displaystyle\sum\limits_{\mu^{\prime} = \pm 1}
      C_{l_{f},l_{ph},l_{ph}+1}^{m \mu^{\prime}} \: f_{l_{f},l_{ph}+1}^{m \mu^{\prime}}(\theta),
\end{array}
\label{eq.2.5.4}
\end{equation}
where
\begin{equation}
\begin{array}{lcl}
  \vspace{2mm}
  C_{l_{f} l_{ph} n}^{m \mu^{\prime}} & = &
    (-1)^{l_{f}+n+1 - \mu^{\prime} + \frac{|m+\mu^{\prime}|}{2}} \;
    (n, 1, l_{ph} \big| -m-\mu^{\prime}, \mu^{\prime}, -m) \; \times \\
  \vspace{2mm}
  & \times &
    \sqrt{\displaystyle\frac{(2l_{f}+1)\,(2n+1)}{32\pi}\;
          \displaystyle\frac{(l_{f}-1)!}{(l_{f}+1)!} \;
          \displaystyle\frac{(n-|m+\mu^{\prime}|)!}{(n+|m+\mu^{\prime}|)!}}, \\
  f_{l_{f} n}^{m \mu^{\prime}}(\theta) & = &
    P_{l_{f}}^{1}(\cos{\theta}) \; P_{1}^{1}(\cos{\theta}) \; P_{n}^{|m+\mu^{\prime}|} (\cos{\theta}).
\end{array}
\label{eq.2.5.5}
\end{equation}
One can see that integration of functions (\ref{eq.2.5.4}) by angle $\theta$ with limits from 0 to $\pi$ gives the total matrix elements $p_{l}^{M}$ and $p_{l}^{E}$ exactly for transition into superposition of all possible final states with different $m_{f}$ at the same $l_{f}$.



We shall find the matrix element at the first values of $l_{f}$ and $l_{ph}$. We have $l_{f} = 1$, $l_{ph} = 1$. Calculating coefficients $C_{11 n}^{m \mu^{\prime}}$ and
functions $f_{1 n}^{m \mu^{\prime}}(\theta)$,
from eq.~(\ref{eq.2.5.4}) we obtain:
\begin{equation}
\begin{array}{lcl}
  \vspace{2mm}
  \displaystyle\frac{d \,\tilde{p}_{1}^{M}}{\sin{\theta}\,d\theta} & = &
    - \displaystyle\frac{3}{8} \: \sqrt{\displaystyle\frac{1}{\pi}} \cdot
    J(1,1) \cdot
    \sin^{2}{\theta} \cos{\theta}, \\
  \vspace{2mm}
  \displaystyle\frac{d \,\tilde{p}_{1}^{E}}{\sin{\theta}\,d\theta} & = &
    i \: \displaystyle\frac{1}{8\sqrt{\pi}}\:
    \sin^{2}{\theta}\:
    \Bigl\{ \sqrt{2} \cdot J(1,0) +  J(1,2) \cdot \Bigl( 1 - 3 \sin^{2}{\theta} \Bigr) \Bigr\}.
\end{array}
\label{eq.2.5.6}
\end{equation}
Integrating these expressions over angle $\theta$, we find the integral matrix elements:
\begin{equation}
\begin{array}{lcllcl}
  \tilde{p}_{1}^{M} & = & 0, &
  \tilde{p}_{1}^{E} & = &
    i \: \displaystyle\frac{1}{6} \, \sqrt{\displaystyle\frac{2}{\pi}} \cdot
    \Bigl\{ J(1,0) - \displaystyle\frac{7}{10} \, \sqrt{2} \cdot J(1,2) \Bigr\}.
\end{array}
\label{eq.2.5.7}
\end{equation}

\subsection{Angular probability of emission of photon with impulse $\mathbf{k}$ and polarization $\mathbf{e}^{(\alpha)}$
\label{sec.2.6}}

I define the probability of transition of the system for time unit from the initial $i$-state into the final $f$-states, being in the given interval $d \nu_{f}$, with emission of photon with possible impulses inside the given interval $d \nu_{ph}$, so
(see Ref.~\cite{Landau.v3.1989}, (42,5) \S~42, p.~189; Ref.~\cite{Berestetsky.1989}, \S~44, p.~191):
\begin{equation}
\begin{array}{l}
  \vspace{2mm}
  d W = \displaystyle\frac{|a_{fi}|^{2}}{T} \cdot d\nu =
    2\pi \:|F_{fi}|^{2} \: \delta (w_{f} - w_{i} + w) \cdot d\nu, \\
  \begin{array}{ll}
    d \nu = d\nu_{f} \cdot d\nu_{ph}, &
    d \nu_{ph} = \displaystyle\frac{d^{3} k}{(2\pi)^{3}} =
          \displaystyle\frac{w^{2} \, dw \,d\Omega_{ph}}{(2\pi c)^{3}},
  \end{array}
\end{array}
\label{eq.2.6.1}
\end{equation}
where $d\nu_{ph}$ and $d\nu_{f}$ are intervals defined for photon and particle in the final $f$-state,
$d\Omega_{ph} = d\,\cos{\theta_{ph}} = \sin{\theta_{ph}} \,d\theta_{ph} \,d\varphi_{ph}$, $k_{ph}=w/c$.
$F_{fi}$ is integral over space with summation by quantum numbers of the system in the final $f$-state. Such procedure is averaging by these characteristics and, so, $F_{fi}$ is independent on them. Interval $d\,\nu_{f}$ has only new characteristics and quantum numbers, by which integration and summation in $F_{fi}$ was not performed.
Integrating eq.~(\ref{eq.2.6.1}) over $dw$ and substituting eq.~(\ref{eq.2.1.2}), we find:
\begin{equation}
\begin{array}{ll}
  \vspace{2mm}
  d W = \displaystyle\frac{Z_{eff}^{2} \,e^{2}}{m^{2}}\:
        \displaystyle\frac{\hbar\, w_{fi}}{2\pi \,c^{3}} \; \Bigl|p(k_{i}, k_{f})\Bigr|^{2} \;
        d \Omega_{ph} \, d\nu_{f}, &
  w_{fi} = w_{i} - w_{f} = \displaystyle\frac{E_{i} - E_{f}}{\hbar}.
\end{array}
\label{eq.2.6.2}
\end{equation}
This is the probability of the photon emission with impulse $\mathbf{k}$ (and with averaging by polarization $\mathbf{e}^{(\alpha)}$) where the integration over angles of the particle motion after the photon emission has already fulfilled.

I define the following probability of emission of photon with momentum $\mathbf{k}_{ph}$ when after such emission the particle moves (or tunnels) along direction $\mathbf{n}_{r}^{f}$: \emph{differential probability concerning angle $\theta$ is such a function, definite integral of which over the angle $\theta$ with limits from 0 to $\pi$ equals to the total probability of the photon emission (\ref{eq.2.6.2})}.
Let us consider function:
\begin{equation}
\begin{array}{ccl}
  \displaystyle\frac{d W(\theta_{f})} {d\,\Omega_{ph} \: d\cos{\theta_{f}}} & = &
  \displaystyle\frac{Z_{eff}^{2}\, \hbar\, e^{2}}{2\pi\, c^{3}}\: \displaystyle\frac{w_{fi}}{m^{2}} \;
    \biggl\{p\,(k_{i},k_{f}) \displaystyle\frac{d\, p^{*}(k_{i},k_{f}, \theta_{f})}{d\cos{\theta_{f}}} + {\rm h. e.} \biggr\}.
\end{array}
\label{eq.2.6.3}
\end{equation}
This probability is inversely proportional to normalized volume $V$. With a purpose to have the probability independent on $V$, I divide eq.~(\ref{eq.2.6.3}) on flux $j$ of outgoing $\alpha$-particles, which is inversely proportional to this volume $V$ also. Using quantum field theory approach
(where $v(\mathbf{p}) = |\mathbf{p}| / p_{0}$ at $c=1$, see~\cite{Bogoliubov.1980}, \S~21.4, p.~174):
\begin{equation}
\begin{array}{cc}
  j = n_{i}\, v(\mathbf{p}_{i}), &
  v_{i} = |\mathbf{v}_{i}| = \displaystyle\frac{c^{2}\,|\mathbf{p}_{i}|} {E_{i}} =
          \displaystyle\frac{\hbar\,c^{2}\,k_{i}} {E_{i}},
\end{array}
\label{eq.2.6.4}
\end{equation}
where $n_{i}$ is average number of particles in time unit before photon emission (we have $n_{i}=1$ for the normalized wave function in the initial $i$-state), $v(\mathbf{p}_{i})$ is module of velocity of outgoing particle in the frame system where colliding center is not moved, I obtain the \emph{differential absolute probability}
(while let's name $dW$ as the \emph{relative probability}):
\begin{equation}
\begin{array}{ccl}
  \vspace{2mm}
  \displaystyle\frac{d\,P (\varphi_{f}, \theta_{f})}{d\Omega_{ph}\, d\cos{\theta_{f}}} & = &
  \displaystyle\frac{d\,W (\varphi_{f}, \theta_{f})}{d\Omega_{ph}\, d\cos{\theta_{f}}} \cdot
    \displaystyle\frac{E_{i}} {\hbar\, c^{2}\, k_{i}} = \\
  & = &
    \displaystyle\frac{Z_{eff}^{2} \,e^{2}}{2\pi\,c^{5}}\:
      \displaystyle\frac{w_{ph}\,E_{i}}{m^{2}\,k_{i}} \;
      \biggl\{p\,(k_{i},k_{f}) \displaystyle\frac{d\, p^{*}(k_{i},k_{f}, \Omega_{f})}{d\,\cos{\theta_{f}}} + {\rm h. e.} \biggr\}.
\end{array}
\label{eq.2.6.5}
\end{equation}
Note that alternative theoretical way for calculations of the angular bremsstrahlung probabilities in $\alpha$-decays was developed in \cite{Jentschura.2008.PRC} based on different definition of the angular probability, different connection of the matrix element with the angle $\theta$ between fragment and photon emitted, application of some approximations.



Let us find the bremsstrahlung probability at the first values $l_{i}=0$, $l_{f}=1$ and $l_{ph}=1$. Starting from eqs.~(\ref{eq.2.3.5}) and (\ref{eq.2.3.6}), and using the found differential and integral electrical and magnetic components (\ref{eq.2.5.6}) and (\ref{eq.2.5.7}),
I calculate:
\begin{equation}
\begin{array}{lcl}
\vspace{2mm}
  \tilde{p}_{1}\, (k_{i},k_{f}) & = &
    - i \, \sqrt{\displaystyle\frac{1}{3}} \cdot
    \Bigl\{ J(1,0) - \displaystyle\frac{7}{10} \, \sqrt{2} \cdot J(1,2) \Bigr\}, \\

  \displaystyle\frac{d \, \tilde{p}_{1}\, (k_{i},k_{f})}{\sin{\theta}\,d\theta} & = &
    i\; \displaystyle\frac{\sqrt{6}}{8} \: \cdot
    \biggl\{
      3\,J(1,1) \cdot \cos{\theta} - \sqrt{2}\, J(1,0)\; - \\
  & - & J(1,2) \cdot \Bigl( 1 - 3 \sin^{2}{\theta} \Bigr)
    \biggr\} \cdot \sin^{2}{\theta}
\end{array}
\label{eq.2.7.1}
\end{equation}
and from eq.~(\ref{eq.2.6.5}) I obtain the \emph{absolute} angular probability:
\begin{equation}
\begin{array}{ccl}
\vspace{3mm}
  \displaystyle\frac{d P^{E1+M1}_{1}(\theta_{f})} {d\,\Omega_{ph} \: d\cos{\theta_{f}}} & = &
    \displaystyle\frac{Z_{eff}^{2}\, e^{2}}{8\,\pi\, c^{5}}\:
    \displaystyle\frac{w_{fi}}{m^{2}}\,
    \displaystyle\frac{E_{i}}{k_{i}}\;
    \biggl\{ \Bigl[ J(1,0) - \displaystyle\frac{7}{10} \, \sqrt{2} \cdot J(1,2) \Bigr]\; \times \\
  & \times &
      \Bigl[
        J^{*}(1,0) +
        \displaystyle\frac{1}{\sqrt{2}} J^{*}(1,2) \cdot \Bigl( 1 - 3 \sin^{2}{\theta} \Bigr)\; - \\
  & - &
    \displaystyle\frac{3}{\sqrt{2}} \,J^{*}(1,1) \cdot \cos{\theta} \Bigr]
    + {\rm h. e.} \biggr\} \cdot \sin^{2}{\theta}.
\end{array}
\label{eq.2.7.2}
\end{equation}

\subsection{Proton--nucleus potential
\label{sec.2.8}}

In order to describe the proton-decay, we need to determine the wave function of a nucleus consisting of a single proton interacting with a residual core.
To describe the proton interacting with the spherically symmetric core we shall use the single-particle potential
in standard optical model
\begin{equation}
  V (r) = v_{c} (r) + v_{N} (r) + v_{\rm so} (r) + v_{l} (r),
\label{eq.2.8.1}
\end{equation}
where $v_{c} (r)$, $v_{N} (r)$, $v_{\rm so} (r)$ and $v_{l} (r)$ are Coulomb, nuclear, spin-orbital and centrifugal components given form in famous paper~\cite{Becchetti.1969.PR}
(see eqs.~(5) in the cited paper) as
\begin{equation}
\begin{array}{lll}
  \vspace{1mm}
  v_{N} (r) = - \displaystyle\frac{V_{R}} {1 + \exp{\displaystyle\frac{r-R_{R}} {a_{R}}}},
  \hspace{2mm}
  v_{l} (r) = \displaystyle\frac{l\,(l+1)} {2mr^{2}}, \\
  \vspace{1mm}
  v_{\rm so} (r) =
    V_{\rm so}\,
    {\mathbf {q \cdot l}}\,
    \displaystyle\frac{\lambda_{\pi}^{2}}{r}\,
    \displaystyle\frac{d}{dr}\, \Bigl[1 + \exp\Bigl(\displaystyle\frac{r-R_{\rm so}} {a_{\rm so}} \Bigr)\Bigr]^{-1}, \\
  v_{c} (r) =
  \left\{
  \begin{array}{ll}
    \displaystyle\frac{Z e^{2}} {r}, &
      \mbox{at  } r \ge R_{c}, \\
    \displaystyle\frac{Z e^{2}} {2 R_{c}}\;
      \biggl\{ 3 -  \displaystyle\frac{r^{2}}{R_{c}^{2}} \biggr\}, &
      \mbox{at  } r < R_{c}.
  \end{array}
  \right.
\end{array}
\label{eq.2.8.2}
\end{equation}
Here, $V_{R}$ and $V_{\rm so}$ are strength of nuclear and spin-orbital components defined in MeV as
(see eq.~(8) in~\cite{Becchetti.1969.PR})
\begin{equation}
\begin{array}{ll}
  V_{R} = 54.0 - 0.32\,E + 0.4\,Z/A^{1/3} + 24.0\,I, &
  V_{\rm so} =6.2,
\end{array}
\label{eq.2.8.3}
\end{equation}
where $I = (N-Z)/A$, $A$ and $Z$ are mass and proton numbers of the daughter nucleus, $E$ is incident lab energy.
$R_{c}$ and $R_{R}$ are Coulomb and nuclear radiuses of nucleus, $a_{R}$ and $a_{\rm so}$ are diffusion parameters.
In calculations I use (all parameters are defined in~\cite{Becchetti.1969.PR}):
\begin{equation}
\begin{array}{llll}
  R_{R} = r_{R}\, A^{1/3}, &
  R_{c} = r_{c}\, A^{1/3}, &
  R_{\rm so} = r_{\rm so}\, A^{1/3}, &
  a_{R} = 0.75\; {\rm fm}, \\
  r_{R} = 1.17\;{\rm fm}, &
  r_{c} = 1.22\;{\rm fm}, &
  r_{\rm so} = 1.01\;{\rm fm}, &
  a_{\rm so} = 0.75\; {\rm fm}.
\end{array}
\label{eq.2.8.4}
\end{equation}

\section{Results
\label{sec.results}}



Let us estimate the bremsstrahlung probability accompanying the proton-decay.
I calculate the bremsstrahlung probability by eq.~(\ref{eq.2.7.2}) for $l_{i}=0$ or by eq.~(\ref{eq.2.6.5}) for $l_{i} \ne 0$.
The potential of interaction between the proton and the daughter nucleus is defined in eqs.~(\ref{eq.2.8.1})--(\ref{eq.2.8.2}) with parameters calculated by eqs.~(\ref{eq.2.8.3})--(\ref{eq.2.8.4}).
To choose the convenient proton-emitters for calculations and analysis, I used systematics presented in Ref.~\cite{Aberg_Nazarewicz.1997.PRC} (see Table II in the cited paper) and selected the $^{157}{\rm Ta}$, $^{161}{\rm Re}$, $^{167}{\rm Ir}$ and $^{185}{\rm Bi}$ nuclei decaying from the $2s_{1/2}$ state (i.e. at $l_{i}=0$), the $^{109}_{53}{\rm I}_{56}$, $^{112}_{55}{\rm Cs}_{57}$ nuclei decaying from the $1d_{5/2}$ state and
the $^{146}_{69}{\rm Tm}_{77}$, $^{151}_{71}{\rm Lu}_{80}$ nuclei decaying from the $0h_{11/2}$ state
(in order to analyze practical effectiveness of the model and convergence of calculations at $l_{i} \ne 0$).
The wave functions of the decaying system in the states before and after the photon emission are calculated
concerning such potential in the spherically symmetric approximation.

\begin{figure}[htbp]
\centerline{\includegraphics[width=88mm]{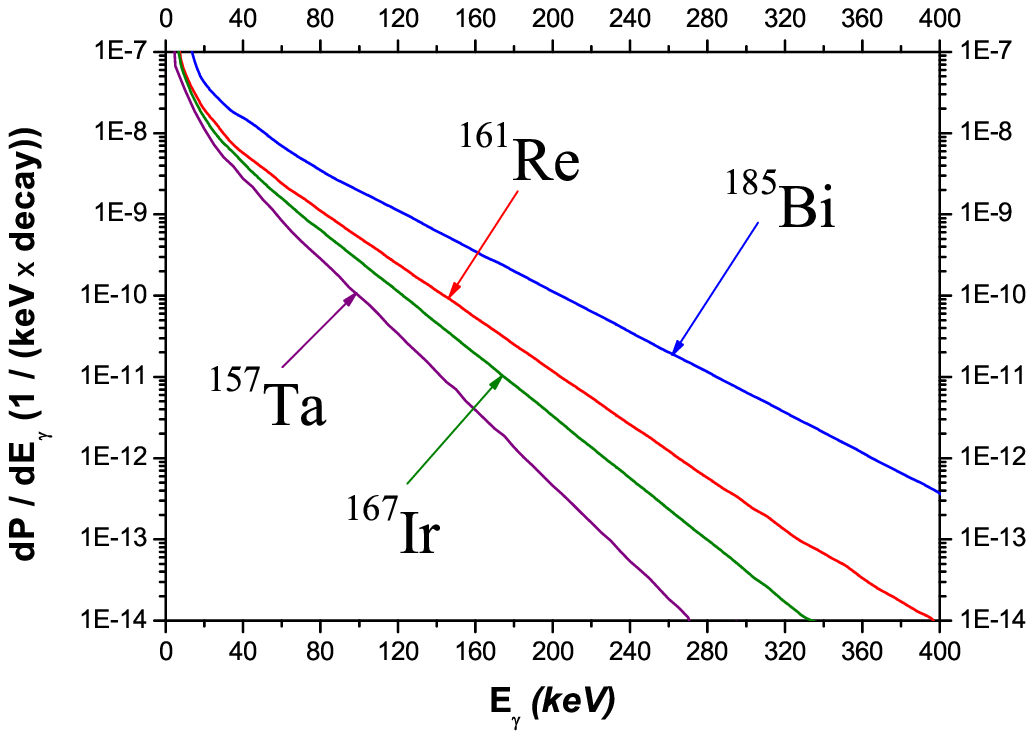}
\includegraphics[width=88mm]{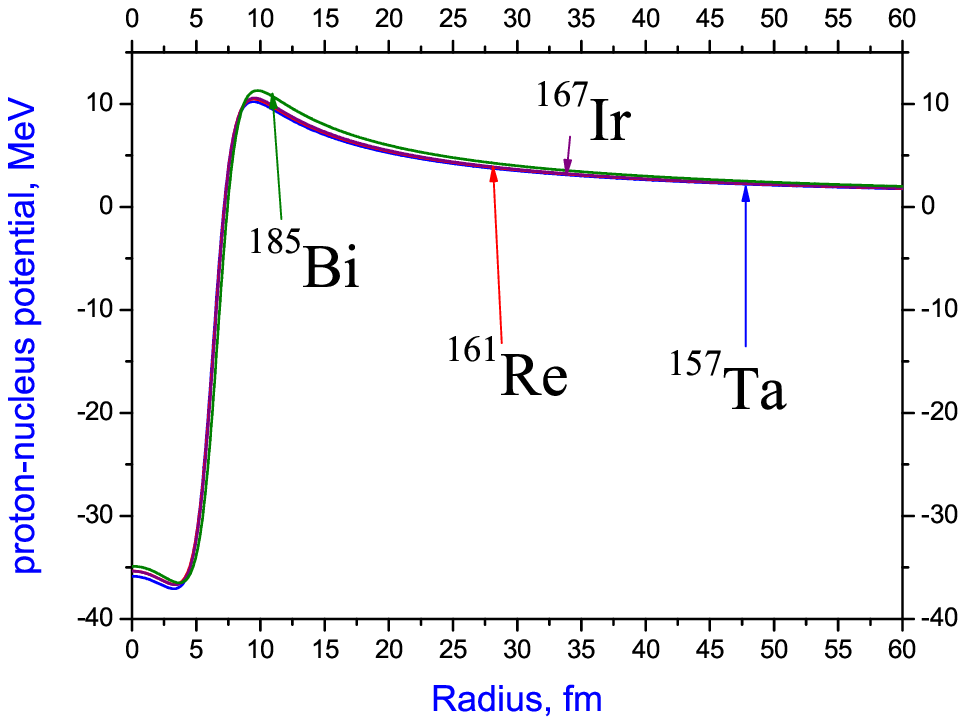}}
\vspace{-8mm}
\caption{\small
The bremsstrahlung during proton decay of the $^{157}{\rm Ta}$, $^{161}{\rm Re}$, $^{167}{\rm Ir}$ and $^{185}{\rm Bi}$ nuclei decaying from the $2s_{1/2}$ state (at the chosen angle $\theta=90^{\circ}$):
(a) the absolute bremsstrahlung probabilities for these nuclei have the different trends that is explained by different tunneling regions for them,
(b) the proton-nucleus potentials for these nuclei are similar practically.
\label{fig.1}}
\end{figure}

The result of calculations of the bremsstrahlung probabilities during proton decay of the nuclei pointed out above from from the initial state $1s_{1/2}$ (at the chosen angle $\theta=90^{\circ}$ between the directions of the proton motion (with its possible tunneling) and the photon emission)
are presented in Fig.~\ref{fig.1}.
In Table~\ref{table.1} one can see some parameters for such studied nuclei. From here one can find that the different proton-emitters have practically similar effective charges, but different essentially tunneling regions. These difference between tunneling regions explain difference between the bremsstrahlung probabilities for such nuclei
(that one can see in Fig.~\ref{fig.1}~(a)).
\begin{table}
\begin{center}
\begin{tabular}{|c|c|c|c|c|c|c|c|c|} \hline
 \multicolumn{3}{|c|}{Proton decay data} &
 \multicolumn{2}{|c|}{Parameters} &
 \multicolumn{2}{|c|}{Turning points} &
 Tunneling &
 \\ \cline{1-7}
  Nucleus &
  $Q_{p}$, keV &
  $T_{1/2}^{\rm WKB}$, sec &
  $R_{R}$, fm & $V_{R}$, MeV &
  1-st, fm & 2-nd, fm &
  region, fm &
  $E_{\rm eff}^{2}$
  \\ \hline
  $^{157}_{73}{\rm Ta}_{83}$ & 947  & 210 $ms$
      & 6.29 & -60.89 & 7.25 & 110.96 & 103.71 & 0.286259 \\
  $^{161}_{75}{\rm Re}_{86}$ & 1214 & 180 $\mu s$
      & 6.3517 & -60.8638 & 7.32 & 88.96 & 81.64 & 0.285328 \\
  $^{167}_{77}{\rm Ir}_{90}$ & 1086 & 35 $ms$
      & 6.3912 & -61.0585 & 7.32 & 100.79 & 93.46 & 0.287924 \\
  $^{185}_{83}{\rm Bi}_{98}$ & 1611 & 3.1 $\mu s$
      & 6.6546 & -61.8599 & 7.56 & 74.28 & 66.72 & 0.303988 \\
  \hline
\end{tabular}
\end{center}
\caption{Parameters of the proton emitters decaying from the $2s_{1/2}$ state ($E_{\rm eff}^{2}$ is square of effective charge): one can see that these nuclei have similar effective charges but different $Q$-values and tunneling regions. This explains the different trends of the spectra presented in Fig.~\ref{fig.1}~(a).
\label{table.1}}
\end{table}

Calculations of the bremsstrahlung probabilities for the case of $l_{i} \ne 0$ are essentially more complicated. Results of such calculations for the $^{109}_{53}{\rm I}_{56}$, $^{112}_{55}{\rm Cs}_{57}$ nuclei decaying from the $1d_{5/2}$ state and the $^{146}_{69}{\rm Tm}_{77}$, $^{151}_{71}{\rm Lu}_{80}$ nuclei decaying from the $0h_{11/2}$ state are presented in next Fig.~\ref{fig.2} (at angle $\theta=90^{\circ}$).
One can see that these spectra have similar orders as results for the previous case at $l_{i}=0$.
\begin{figure}[htbp]
\centerline{\includegraphics[width=88mm]{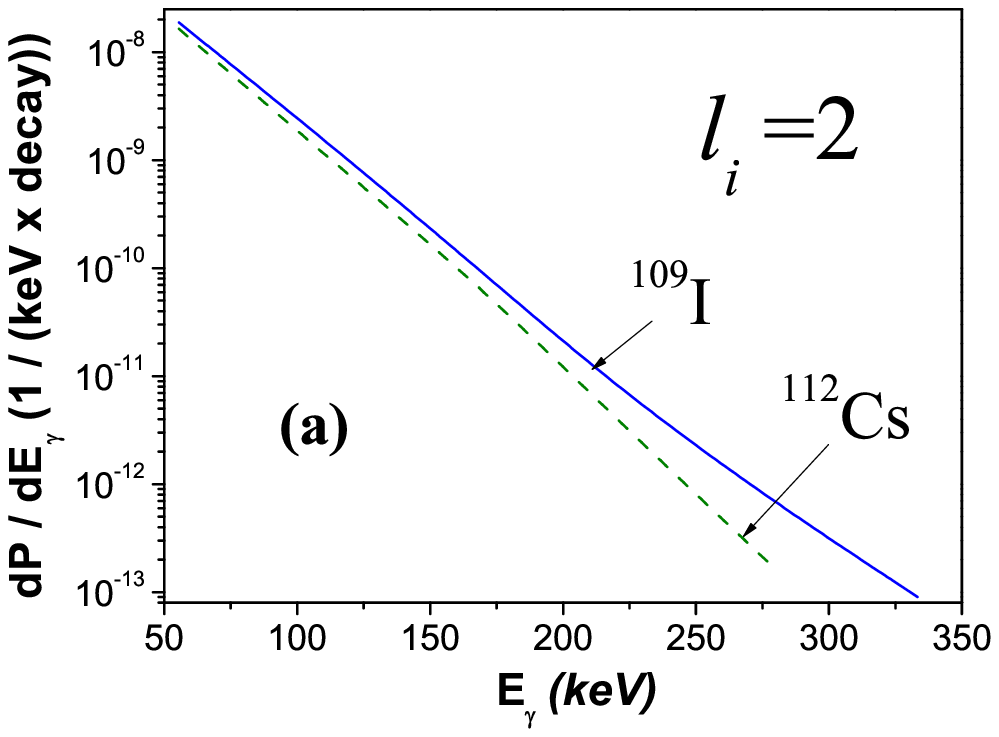}
\hspace{-5mm}\includegraphics[width=88mm]{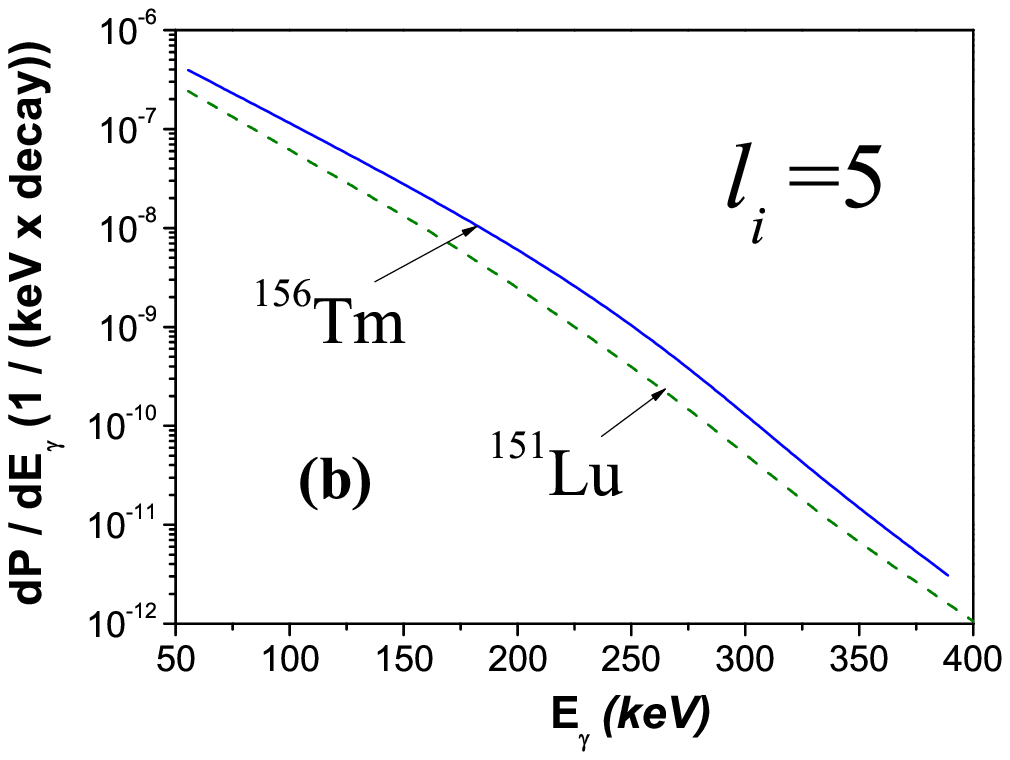}}
\vspace{-8mm}
\caption{\small
(a) The bremsstrahlung accompanying proton decay of the $^{109}{\rm I}$ and $^{112}{\rm Cs}$ nuclei from the state $1d_{5/2}$ (at $\theta=90^{\circ}$);
(b) the bremsstrahlung accompanying proton decay of the $^{146}{\rm Tm}$ and $^{151}{\rm Lu}$ nuclei from the state $0h_{11/2}$ (at $\theta=90^{\circ}$).
\label{fig.2}}
\end{figure}
According to analysis, inclusion of the spin-orbital component $v_{\rm so}$ into the total potential (\ref{eq.2.8.1}) changes the resulting probability less than 1 percent (practically, 3-rd or 4-th digit in spectrum is changed; such spin-orbital influence on the spectrum has been estimated for the first time in the tasks of the bremsstrahlung during different types of nuclear decays).
In next Fig.~\ref{fig.3} angular distribution of the bremsstrahlung probability is presented.
In particular, one can see that we obtain convergent calculations of angular integrals and the spectra at $l_{i} \ne 0$.
\begin{figure}[htbp]
\centerline{\includegraphics[width=88mm]{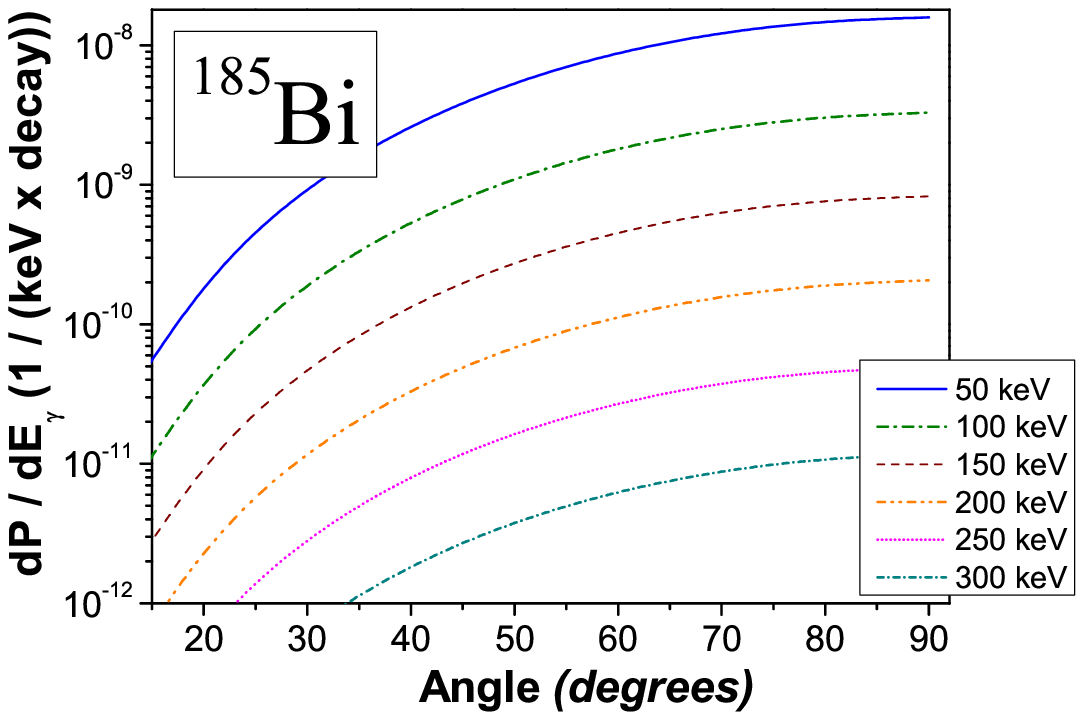}
\hspace{0mm}\includegraphics[width=88mm]{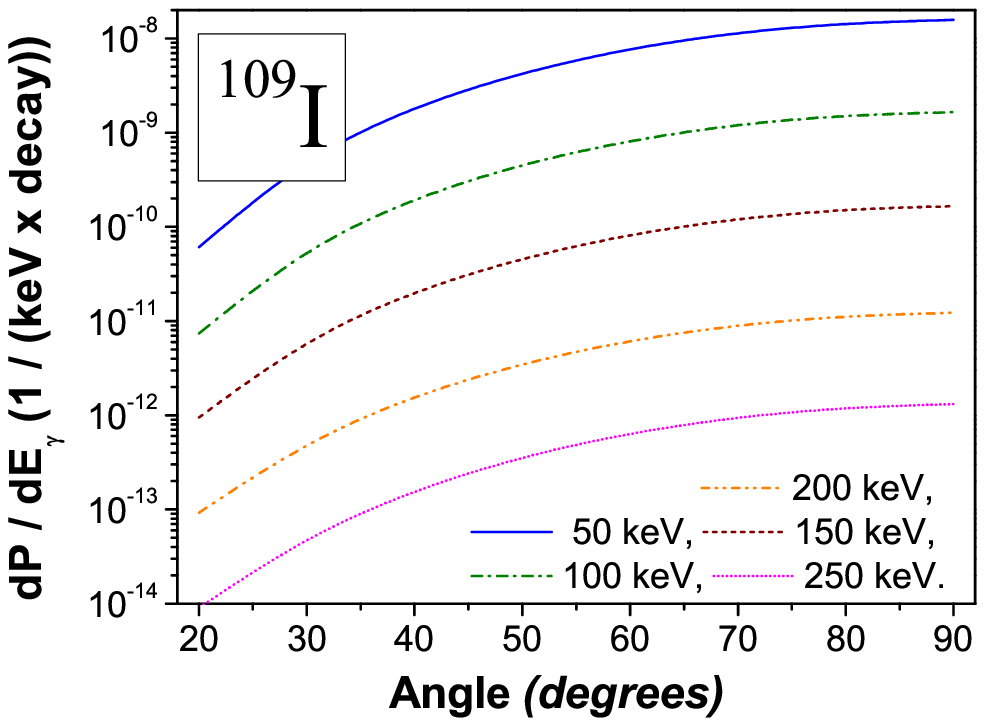}}
\vspace{-8mm}
\caption{\small
The bremsstrahlung probability during proton decay of the $^{185}{\rm Bi}$ (from the state $2s_{1/2}$) and $^{109}{\rm I}$ (from the state $1d_{5/2}$) nuclei in dependence on the $\theta$ angle between direction of motion (with possible tunneling) of proton and direction of the emission of photon.
In these figures different curves correspond to different energies of energy of the photon emitted.
\label{fig.3}}
\end{figure}


\section{Conclusion
\label{sec.conclusions}}

In the paper the emission of the bremsstrahlung photons accompanying the proton decay of nuclei, is studied (for the first time, in systematic basis). The new improved multipolar model describing such a process is presented. The angular formalism of calculations of the matrix elements is stated in details. The absolute bremsstrahlung probabilities for the $^{157}{\rm Ta}$, $^{161}{\rm Re}$, $^{167}{\rm Ir}$ and $^{185}{\rm Bi}$ nuclei decaying from the $2s_{1/2}$ state (i.e. at $l_{i}=0$), the $^{109}_{53}{\rm I}_{56}$ and $^{112}_{55}{\rm Cs}_{57}$ nuclei decaying from the $1d_{5/2}$ state (i.e. at $l_{i} = 2$), the $^{146}_{69}{\rm Tm}_{77}$ and $^{151}_{71}{\rm Lu}_{80}$ nuclei decaying from the $0h_{11/2}$ state (i.e. at $l_{i}=5$) are predicted. Such spectra have orders of values similar to the experimental spectra of the bremsstrahlung photons emitted during the $\alpha$-decay (see \cite{Giardina.2008.MPLA,Kasagi.1997.JPHGB,Kasagi.1997.PRLTA,Boie.2007.PRL,Maydanyuk.2008.EPJA}).
It needs to note that arsenal of experimental measurements of the bremsstrahlung emission during different types of nuclear decays is not rich. A serious difficulty in obtaining of desirable accuracy in measurements lies in small values of the photon emission probability. One can hope that the presented theoretical estimations of the bremsstrahlung spectra indicate on the new possibility to perform further measurements and
to study bremsstrahlung photons during proton decay experimentally.





\appendix
\section{Angular integrals at arbitrary $l_{i}$
\label{app.6}}

Let us calculate angular integrals:
\begin{equation}
\begin{array}{ccl}
  I_{M}\, (l_{i}, l_{f}, l_{\rm ph}, l_{1}) & = &
    \displaystyle\sum\limits_{\mu = \pm 1}
    \displaystyle\sum\limits_{m_{i}}
    \displaystyle\sum\limits_{m_{f}}
      \mu h_{\mu}
    \displaystyle\int
      Y_{l_{f}m_{f}}^{*}({\mathbf n}_{\rm r}^{f})\,
      \mathbf{T}_{l_{i}\, l_{1},\, m_{i}}(\mathbf{n}^{i}_{\rm r})\,
      \mathbf{T}_{l_{\rm ph}\,l_{\rm ph},\, \mu}^{*}({\mathbf n}_{\rm ph})\; d\Omega, \\

  I_{E}\, (l_{i}, l_{f}, l_{\rm ph}, l_{1}, l_{2}) & = &
    \displaystyle\sum\limits_{\mu = \pm 1}
    \displaystyle\sum\limits_{m_{i}}
    \displaystyle\sum\limits_{m_{f}}
      h_{\mu}
    \displaystyle\int
      Y_{l_{f}m_{f}}^{*}({\mathbf n}_{\rm r}^{f})\,
      \mathbf{T}_{l_{i} l_{1},\, m_{i}}(\mathbf{n}^{i}_{\rm r})\,
      \mathbf{T}_{l_{\rm ph} l_{2},\, \mu}^{*}({\mathbf n}_{\rm ph})\; d\Omega,
\end{array}
\label{eq.app.6.1}
\end{equation}
%
%
%
%
Substituting the functions $\mathbf{T}_{jl,m}(\mathbf{n})$ in form (\ref{eq.2.3.3}), we obtain:
\begin{equation}
\begin{array}{l}
\vspace{1mm}
  I_{M}\, (l_{i}, l_{f}, l_{\rm ph}, l_{1}) =
    \displaystyle\sum\limits_{\mu = \pm 1}
    \displaystyle\sum\limits_{m_{i}m_{f}}
      \mu h_{\mu}
      \displaystyle\sum\limits_{\mu^{\prime} = \pm 1}
        (l_{1}, 1, l_{i} \,\big| \,m_{i}-\mu^{\prime}, \mu^{\prime}, m_{i})\; \times \\
\vspace{3mm}
  \times\;
        (l_{\rm ph}, 1, l_{\rm ph} \,\big|\, \mu-\mu^{\prime}, \mu^{\prime}, \mu)\,
    \displaystyle\int
      Y_{l_{f}m}^{*}({\mathbf n}_{\rm r}^{f}) \cdot
      Y_{l_{1},\, m_{i}-\mu^{\prime}}(\mathbf{n}^{i}_{\rm r}) \cdot
      Y_{l_{\rm ph},\, \mu-\mu^{\prime}}^{*} (\mathbf{n}_{\rm ph})\; d\Omega, \\
\vspace{1mm}
  I_{E}\, (l_{i}, l_{f}, l_{\rm ph}, l_{1}, l_{2}) =
    \displaystyle\sum\limits_{\mu = \pm 1}
    \displaystyle\sum\limits_{m_{i}m_{f}}
      h_{\mu}
      \displaystyle\sum\limits_{\mu^{\prime} = \pm 1}
        (l_{1}, 1, l_{i} \,\big| \,m_{i}-\mu^{\prime}, \mu^{\prime}, m_{i}) \\
  \times
        (l_{2}, 1, l_{\rm ph} \,\big|\, \mu-\mu^{\prime}, \mu^{\prime}, \mu)\;
    \displaystyle\int
      Y_{l_{f}m}^{*}({\mathbf n}_{\rm r}^{f}) \cdot
      Y_{l_{1},\, m_{i}-\mu^{\prime}} (\mathbf{n}^{i}_{\rm r}) \cdot
      Y_{l_{2},\, \mu-\mu^{\prime}}^{*} (\mathbf{n}_{\rm ph})\; d\Omega.
\end{array}
\label{eq.app.6.3}
\end{equation}
Here we have used the orthogonality condition for vectors $\xi_{\pm 1}$ ($\xi_{0} = 0$).

Taking into account different variants of definition of spherical functions $Y_{lm}(\theta, \varphi)$,
we define them according to \cite{Landau.v3.1989} (see. p.~119, (28,7)--(28,8)):
%
\begin{equation}
\begin{array}{lcl}
  Y_{lm}(\theta,\varphi) & = &
    (-1)^{\frac{m+|m|}{2}} \; i^{l} \;
    \sqrt{\displaystyle\frac{2l+1}{4\pi} \displaystyle\frac{(l-|m|)!}{(l+|m|)!} }  \; P_{l}^{|m|} (\cos{\theta}) \cdot e^{im\varphi},
\end{array}
\label{eq.app.6.4}
\end{equation}
where $P_{l}^{m}(\cos{\theta})$ are \emph{associated Legandre's polynomials}. 
Rewrite the angular integral in (\ref{eq.app.6.3}) as
\begin{equation}
\begin{array}{l}
  \displaystyle\int \:
    Y_{l_{f} m_{f}}^{*}({\mathbf n}_{\rm r}) \,
    Y_{l_{1},\, m_{i}-\mu^{\prime}}(\mathbf{n}_{\rm r}) \,
    Y_{n,\, \mu-\mu^{\prime}}^{*}(\mathbf{n}_{\rm r}) \;
    d\Omega = \\


  = (-1)^{\frac{m_{f}+|m_{f}| + m_{i} - \mu^{\prime} + |m_{i} - \mu^{\prime}| +
            \mu-\mu^{\prime}+|\mu-\mu^{\prime}|}{2}}\; (-1)^{l_{f}+n} \; i^{l_{f}+l_{1}+n} \times \\
 \; \times
    \sqrt{\displaystyle\frac{2l_{f}+1}{4\pi} \displaystyle\frac{(l_{f}-|m_{f}|)!}{(l_{f}+|m_{f}|)!}}\;
    \sqrt{\displaystyle\frac{2l_{1}+1}{4\pi}
      \displaystyle\frac{(l_{1}-|m_{i}-\mu^{\prime}|)!}{(l_{1}+|m_{i}-\mu^{\prime}|)!}}\;
    \sqrt{\displaystyle\frac{2n+1}{4\pi} \displaystyle\frac{(n-|\mu-\mu^{\prime}|)!}{(n+|\mu-\mu^{\prime}|)!}}\;
    \times \\
  \;\times
    \displaystyle\int\limits_{0}^{2\pi}\,
      e^{i(-m_{f}+m_{i}-\mu^{\prime} -\mu+\mu^{\prime})\varphi}\: d\varphi \cdot
    \displaystyle\int\limits_{0}^{\pi}\:
      P_{l_{f}}^{|m_{f}|}(\cos{\theta})\;
      P_{l_{1}}^{|m_{i} - \mu^{\prime}|}(\cos{\theta})\;
      P_{n}^{|\mu-\mu^{\prime}|} (\cos{\theta}) \cdot
      \sin{\theta}\, d\theta.
\end{array}
\label{eq.app.6.5}
\end{equation}
Integral over $\varphi$ in this expression is different from zero only in case
\begin{equation}
\begin{array}{c}
  \mu = m_{i} - m_{f}.
\end{array}
\label{eq.app.6.6}
\end{equation}
So, we obtain the following restrictions:
\begin{equation}
\begin{array}{cc}
  n \ge |\mu - \mu^{\prime}| = |m_{i} - m_{f} + \mu^{\prime}|, & \mu = \pm 1 .
\end{array}
\label{eq.app.6.7}
\end{equation}
Using it, we find:
\begin{equation}
\begin{array}{l}
\vspace{1mm}
  \displaystyle\int \:
    Y_{l_{f}m_{f}}^{*}({\mathbf n}_{\rm r})\,
    Y_{l_{1},\, m_{i}-\mu^{\prime}}(\mathbf{n}_{\rm r})\,
    Y_{n,\, \mu-\mu^{\prime}}^{*}(\mathbf{n}_{\rm r})\;
    d\Omega = \\

%
%

\vspace{1mm}
  = (-1)^{l_{f} + n + m_{i} - \mu^{\prime}}\;
    i^{l_{f}+l_{1}+n + |m_{f}| + |m_{i} - \mu^{\prime}| + |m_{i} - m_{f}-\mu^{\prime}|}\; \times \\
  \;\times
    \sqrt{
      \displaystyle\frac{(2l_{f}+1)\, (2l_{1}+1)\, (2n+1)}{16\pi}
      \displaystyle\frac{(l_{f}-|m_{f}|)!}{(l_{f}+|m_{f}|)!}\;
      \displaystyle\frac{(l_{1}-|m_{i}-\mu^{\prime}|)!} {(l_{1}+|m_{i}-\mu^{\prime}|)!}\;
      \displaystyle\frac{(n-|m_{i} - m_{f}-\mu^{\prime}|)!}{(n+|m_{i} - m_{f} -\mu^{\prime}|)!}}\;
    \times \\

  \;\times
    \displaystyle\int\limits_{0}^{\pi}\:
      P_{l_{f}}^{|m_{f}|}(\cos{\theta})\;
      P_{l_{1}}^{|m_{i} - \mu^{\prime}|}(\cos{\theta})\;
      P_{n}^{|m_{i} - m_{f} - \mu^{\prime}|} (\cos{\theta}) \cdot
      \sin{\theta}\, d\theta.
\end{array}
\label{eq.app.6.8}
\end{equation}

Now we introduce the following generalized coefficient
\begin{equation}
\begin{array}{l}
\vspace{1mm}
  C_{l_{i} l_{f} l_{ph} l_{1} l_{2}}^{m_{i} m_{f} \mu^{\prime}} =
    (-1)^{l_{f} + l_{2} + m_{i} - \mu^{\prime}}\;
    i^{l_{f}+l_{1}+ l_{2} + |m_{f}| + |m_{i} - \mu^{\prime}| + |m_{i} - m_{f}-\mu^{\prime}|}\; \times \\
  \vspace{2mm}
  \times\;
    (l_{1}, 1, l_{i} \,\big| \,m_{i}-\mu^{\prime}, \mu^{\prime}, m_{i})\;
    (l_{2}, 1, l_{\rm ph} \,\big|\, m_{i}-m_{f} -\mu^{\prime}, \mu^{\prime}, m_{i}-m_{f})\; \times \\
  \times\;
    \sqrt{
      \displaystyle\frac{(2l_{f}+1)\, (2l_{1}+1)\, (2l_{2}+1)}{16\pi}
      \displaystyle\frac{(l_{f}-|m_{f}|)!}{(l_{f}+|m_{f}|)!}\;
      \displaystyle\frac{(l_{1}-|m_{i}-\mu^{\prime}|)!}{(l_{1}+|m_{i}-\mu^{\prime}|)!}\;
      \displaystyle\frac{(l_{2}-|m_{i}-m_{f}-\mu^{\prime}|)!} {(l_{2} +|m_{i} - m_{f} -\mu^{\prime}|)!}}
\end{array}
\label{eq.app.6.11}
\end{equation}
and function
\begin{equation}
  f_{l_{f} l_{1} l_{2}}^{m_{i} m_{f} \mu^{\prime}}(\theta) =
    P_{l_{f}}^{|m_{f}|}(\cos{\theta})\;
    P_{l_{1}}^{|m_{i} - \mu^{\prime}|}(\cos{\theta})\;
    P_{l_{2}}^{|m_{i} - m_{f} - \mu^{\prime}|} (\cos{\theta}).
\label{eq.app.6.12}
\end{equation}
In result, we obtain the following expressions for angular integrals $I_{M}$ and $I_{E}$:
\begin{equation}
\begin{array}{l}
  I_{M}\, (l_{i}, l_{f}, l_{\rm ph}, l_{1}) =
    \displaystyle\sum\limits_{m_{i}m_{f}}
    (m_{i}-m_{f}) \,h_{m_{i}-m_{f}} \;
    \displaystyle\sum\limits_{\mu^{\prime} = \pm 1}
      C_{l_{i} l_{f} l_{ph} l_{1} l_{ph}}^{m_{i} m_{f} \mu^{\prime}}
      \displaystyle\int\limits_{0}^{\pi}\:
        f_{l_{f} l_{1} l_{\rm ph}}^{m_{i} m_{f} \mu^{\prime}}(\theta)\; \sin{\theta}\,d\theta, \\

  I_{E}\, (l_{i}, l_{f}, l_{\rm ph}, l_{1}, l_{2}) =
    \displaystyle\sum\limits_{m_{i}m_{f}}
    h_{m_{i}-m_{f}} \;
    \displaystyle\sum\limits_{\mu^{\prime} = \pm 1}
      C_{l_{i} l_{f} l_{ph} l_{1} l_{2}}^{m_{i} m_{f} \mu^{\prime}}
      \displaystyle\int\limits_{0}^{\pi} \:
      f_{l_{f} l_{1} l_{2}}^{m_{i} m_{f} \mu^{\prime}}(\theta)\; \sin{\theta}\,d\theta.
\end{array}
\label{eq.app.6.13}
\end{equation}
We also define differential functions form these angular integrals by angle $\theta$ so:
\begin{equation}
\begin{array}{lcl}
  \displaystyle\frac{d\, I_{M}\, (l_{i}, l_{f}, l_{\rm ph}, l_{1})} {\sin{\theta}\,d\theta} & = &
  \displaystyle\sum\limits_{m_{i}m_{f}}
    (m_{i}-m_{f}) \,h_{m_{i}-m_{f}} \;
    \displaystyle\sum\limits_{\mu^{\prime} = \pm 1}
      C_{l_{i} l_{f} l_{ph} l_{1} l_{ph}}^{m_{i} m_{f} \mu^{\prime}} \cdot
      f_{l_{f} l_{1} l_{\rm ph}}^{m_{i} m_{f} \mu^{\prime}}(\theta), \\

  \displaystyle\frac{d\, I_{E}\, (l_{i}, l_{f}, l_{\rm ph}, l_{1}, l_{2})} {\sin{\theta}\,d\theta} & = &
  \displaystyle\sum\limits_{m_{i}m_{f}}
    h_{m_{i}-m_{f}} \;
    \displaystyle\sum\limits_{\mu^{\prime} = \pm 1}
      C_{l_{i} l_{f} l_{ph} l_{1} l_{2}}^{m_{i} m_{f} \mu^{\prime}} \cdot
      f_{l_{f} l_{1} l_{2}}^{m_{i} m_{f} \mu^{\prime}}(\theta).
\end{array}
\label{eq.app.6.14}
\end{equation}



\end{document}